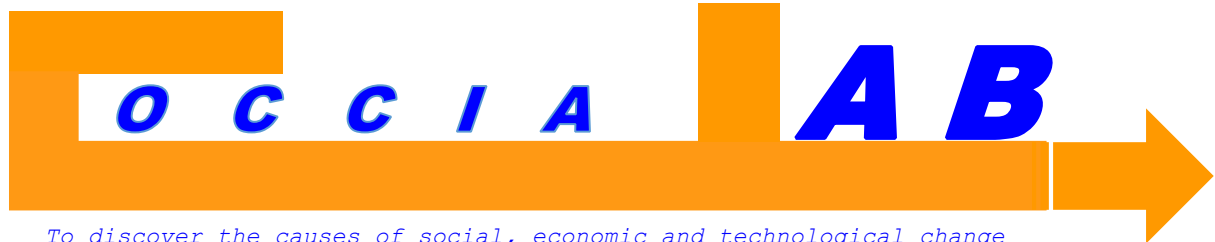

*To discover the causes of social, economic and technological change*

# A New Classification of Technologies

**Mario COCCIA**

ARIZONA STATE UNIVERSITY
CocciaLAB is at the Center for Social Dynamics and Complexity
Interdisciplinary Science and Technology Building 1 (ISBT1)
550 E. Orange Street, Tempe- AZ 85287-4804 USA

and

CNR -- NATIONAL RESEARCH COUNCIL OF ITALY
Via Real Collegio, 30-10024, Moncalieri (TO), Italy

*E*-mail: mario.coccia@cnr.it

# A New Classification of Technologies

*Mario Coccia* [1]

Arizona State University &

CNR -- National Research Council of Italy

*E*-mail: mario.coccia@cnr.it

Mario Coccia ORCID: http://orcid.org/0000-0003-1957-6731

**ABSTRACT.** This study here suggests a classification of technologies based on taxonomic characteristics of interaction between technologies in complex systems that is not a studied research field in economics of technical change. The proposed taxonomy here categorizes technologies in four typologies, in a broad analogy with the ecology: 1) *technological parasitism* is a relationship between two technologies T1 and T2 in a complex system S where one technology T1 benefits from the interaction with T2, whereas T2 has a negative side from interaction with T1; 2) *technological commensalism* is a relationship between two technologies in S where one technology benefits from the other without affecting it; 3) *technological mutualism* is a relationship in which each technology benefits from the activity of the other within complex systems; 4) *technological symbiosis* is a long-term interaction between two (or more) technologies that evolve together in complex systems. This taxonomy systematizes the typologies of interactive technologies within complex systems and predicts their evolutionary pathways that generate stepwise coevolutionary processes of complex systems of technology. This study here begins the process of generalizing, as far as possible, critical typologies of interactive technologies that explain the long-run evolution of technology. The theoretical framework developed here opens the black box of the interaction between technologies that affects, with different types of technologies, the evolutionary pathways of complex systems of technology over time and space. Overall, then, this new theoretical framework may be useful for bringing a new perspective to categorize the gradient of benefit to technologies from interaction with other technologies that can be a ground work for development of more sophisticated concepts to clarify technological and economic change in human society.

---

[1]   I gratefully acknowledge financial support from the CNR - National Research Council of Italy for my visiting at Arizona State University where this research started in 2016 (CNR - NEH Memorandum Grant n. 0072373-2014 and n. 0003005-2016).

## Introduction

Patterns of technological innovation have also been analyzed using analogies with biological phenomena over the last century (Basalla, 1988; Nelson and Winter, 1982; Solé et al., 2013; Sahal, 1981; Veblen, 1904; Wagner, 2011; Ziman, 2000)[2]. Wagner and Rosen (2014) argue that the application of Darwinian and evolutionary biological thinking to different research fields has reduced the distance between life sciences and social sciences generating new approaches, such as the evolutionary theory of economic change (Nelson and Winter, 1982; cf., Dosi, 1988). Basalla (1988) suggests the similarity between history of technology and biological evolution. Usher (1954), within these research fields, analyzed the nature of technological processes and the forces that influenced events at technical level (cf., Ruttan, 2001). In general, technological evolution, as biological evolution, displays radiations, stasis, extinctions, and novelty (Valverde et al., 2007).

Scholars of the economics of technical change have tried of defining, explaining and measuring innovation in its many forms as well as of providing classifications of technical change and progress (Asimakopulos and Weldon, 1963; Bigman, 1979; Coccia, 2006; Freeman and Soete, 1987; Pavitt, 1984; Robinson, 1971)[3]. As a matter of fact, the study and classification of technological innovations are a central and enduring research theme in the economics of technical change (Bowker, 2000; Jones et al., 2012). Although the concepts of "classification" and "taxonomy" are almost synonyms, they have different meaning. The term taxonomy (from ancient Greek word taxon=arrangement, array) refers to a branch of systematics based on the theory and practice of producing classification schemes with the aim of maximizing the differences among groups. Thus, a taxonomic process provides rules on how to form and represent groups with classification. Instead, classification in science is a product of the taxonomic process that represents classes of entities with a matrix, a table, a dendrogram, etc. (McKelvey, 1982). For instance, the biological classification by Linnaeus, the periodic classification of chemical elements by Mendeleev, the Mercalli scale in seismology, the Beaufort wind force scale, etc. (Coccia, 2006). Taxonomy has usefulness in natural and social sciences if it is able to reduce the complexity of the population studied into simple classes, which are represented by a classification (Archibugi, 2001). In particular, social sciences have two general approaches to create a classification: the empirical and theoretical one (Rich, 1992; Doty and Glick, 1994). Theoretical classifications in social sciences begin by developing a theory of differences which then results in a classification of typologies. The empirical approach begins by gathering data about the entities under study. These data are then processed using statistical techniques to produce groups with measures of similarity (e.g., Minkowski distance, Manhattan distance, Euclidean distance, Weighted Euclidean distance, Mahalanobis distance, Chord distance, etc.).

---

2 Cf. also, Wagner (2017) and Weiberger et al. (2017) for process and pattern in innovations from cells to societies.
3 Cf., Rosegger, 1980; Christensen et al., 2015; Coccia, 2005, 2005a; Hall and Rosenberg, 2010.

The subject matter of this study here is taxonomy of technologies. In general, technology studies present several taxonomies of technical change (Coccia, 2006; Freeman and Soete, 1987; Pavitt, 1984). However, a taxonomy that considers the interaction between technologies in complex systems is unknown.

This paper here has two goals. The first is to propose a new taxonomy of technologies based on a taxonomic characteristic of interaction between technologies within complex systems. The second is to explain and generalize, whenever possible this theory that may clarify the typologies of interactive technologies that support paths of technological evolution over time. Overall, then, this theoretical framework here can systematize and predict behavior of interactive technologies and their evolutionary pathways in complex systems, and encourage further theoretical exploration in this *terra incognita* of the interaction between technologies during technological and economic change.

**Theoretical background**

Economics of technical change presents many classifications of technological innovation (Coccia, 2006). De Marchi (2016, p. 983) argues that The Frascati and Oslo manuals assemble technological activities without attempting to propose a cogent organization of the categories. In these research fields, Rosenberg (1982) introduces the distinction between technology directed to new product development, and technology that generates cost reducing–process innovation. Hicks (1932) argued that technological progress is naturally directed to reducing the utilization of a factor that is becoming expansive. Archibugi and Simonetti (1998, pp. 298-299) suggest that each technological innovation can be classified considering:

1. *Technological nature of innovation* that is a technical description of technological innovation. This classification considers the objects of technological change;

2. *The sector of activity of the producing organization.* This is a classification by subject that promotes technological innovation;

3. *The product group where the innovation is used.* Here, it is considered the economic object of technological innovation;

4. *The using organization.* Here too, as in point 2, it is considered the economic subject of technological innovation;

5. *The human needs* which the technological innovation is designed to address.

Freeman and Soete (1987, pp. 55-62, original italics and emphasis) propose a taxonomy to categorize various types of technical change and distinguish:

*Incremental Innovations.* These occur more or less continuously in any industry or service activity, although at a varying rate in different industries and over different time periods. They may often occur, as the outcome of improvements suggested by engineers and others directly engaged in the production process, or as a result of initiatives and proposals by users …. They are particularly important in the follow-through period after a radical breakthrough innovation and frequently associated with the scaling up of plant and equipment and quality improvements to products and services for a variety of specific applications. Although their combined effect is extremely important in the growth of productivity, no single incremental innovation has dramatic effects, and they may sometimes pass unnoticed and unrecorded….

*Radical Innovations.* These are discontinuous events and in recent times is usually the result of a deliberate research and development activity in enterprises and/or in university and government laboratories. They are unevenly distributed over sectors and over time.... big improvements in the cost and quality of existing products .... in terms of their economic impact they are relatively small and localized…. Strictly speaking… radical innovations would constantly require the addition of new rows and columns in an input-output table….

*New Technological Systems.* Keirstead (1948) … introduced the concept of 'constellations' of innovations, which were technically and economically inter-related. Obvious examples are the clusters of synthetic materials innovations and petrochemical innovations in the thirties, forties and fifties…. They include numerous radical and incremental innovations in both products and processes (Freeman et al., 1982).

*Changes of 'Techno-Economic Paradigm' (Technological Revolutions).* These are far-reaching and pervasive changes in technology, affecting many (or even all) branches of the economy, as well as giving rise to entirely new sectors. Examples given by Schumpeter were the steam engine and electric power. Characteristic of this type of technical change is that it affects the input cost structure and the conditions of production and distribution for almost every branch of the economy. A change in techno-economic paradigm thus comprises clusters of radical and incremental innovations and embraces several 'new technological systems'.

Sahal (1985, p. 64, original Italics) argues that technological innovations can be: "*structural innovations* that arise from a process of differential growth; whereby the parts and the whole of a system do not grow at the same rate. Second, we have what may be called the *material innovations* that are necessitated in an attempt to meet the requisite changes in the criteria of technological construction as a consequence of changes in the scale of the object. Finally, we have what may be called the *systems innovations* that arise from integration of two or more symbiotic technologies in an attempt to simplify the outline of the overall structure". This trilogy can generate the emergence of various techniques including revolutionary innovations in a variety of technological and scientific fields (cf., Sahal, 1981; Coccia, 2016, 2016a).

Abernathy and Clark (1985, p. 3) introduce the concept of transilience: "the capacity of an innovation to influence the established systems of production and marketing. Application of the concept results in a categorization of innovation into four types". In particular, the four typologies of innovation by Abernathy and Clark (1985, p. 7ff, original italics) are:

*Architectural innovation.* New technology that departs from established systems of production, and in turn opens up new linkages to markets and users, is characteristic of the creation of new industries as well as the reformation of old ones. Innovation of this sort defines the basic configuration of product and process, and establishes the technical and marketing agendas that will guide subsequent development. In effect, it lays down the architecture of the industry, the broad framework within which competition will occur and develop ….

*Innovation in the market niche* …. Opening new market opportunities through the use of existing technology is central to the kind of innovation that they have labelled "Niche Creation", but here the effect on production and technical systems is to conserve and strengthen established designs …. In some instances, niche creation involves a truly trivial change in technology, in which the impact on productive systems and technical knowledge is incremental. But this type of innovation may also appear in concert with significant new product introductions, vigorous competition on the basis of features, technical refinements, and even technological shifts. The important point is that these changes build on established technical competence, and improve its applicability in emerging market segments ….

*Regular innovation* ….is often almost invisible, yet can have a dramatic cumulative effect on product cost and performance. Regular innovation involves change that builds on established technical and production competence and that is applied to existing markets and customers. The effect of these changes is to entrench existing skills and resources…. can have dramatic effect on production costs, reliability and performance…. Regular innovation can have a significant effect on product characteristics and thus can serve to strengthen and entrench not only competence in production, but linkages to customers and markets….

*Revolution innovation*. Innovation that disrupts and renders established technical and production competence obsolete, yet is applied to existing markets and customers…. The reciprocating engine in aircraft, vacuum tubes, and mechanical calculators are recent examples of established technologies that have been over thrown through a revolutionary design. Yet the classic case of revolutionary innovation is the competitive duel between Ford and GM in the late 1920s and early 1930s.

Anderson and Tushman (1986) distinguish, in patterns of technological innovation, two types of discontinuous change: competence-enhancing and competence-destroying discontinuities. Competence-enhancing discontinuities are based on existing skills and know-how. Competence-destroying discontinuities, instead, require fundamentally new skills and cause obsolescence of existing products and knowledge. In general, technological shifts are due to both competence-destroying and competence-enhancing because some firms can either destroy or enhance the competence existing in industries (*cf.,* Tushman and Anderson, 1986). Usher (1954), in this context, argues that technological innovation is driven by a cumulative significance in the inventive process (cf., Rosenberg, 1982).

Grodal et al. (2015), in management of technology, propose that the evolution of both technological designs and categories follows a similar pattern, characterized by an early period of divergence followed by a period of convergence. Grodal et al. (2015, p. 426) identify the following mechanisms within coevolutionary processes of technology:

- Design recombination is the creative synthesis of two or more previously separate designs that results in the

creation of a new design to address an existing or potential need.

- Path dependence is the mechanism through which the cumulative effects of prior technological design choices increasingly determine and constrain subsequent design recombinations.

- Design competition is the mechanism by which producers and users make design investment choices about which designs to retain and which to abandon.

Garcia and Calantone (2002) apply Boolean logic to identify three labels in product innovation management: radical, really new and incremental innovation. The radical innovations cause discontinuity of marketing and technology, both at a macro and a micro level. Incremental innovations occur only at micro level and cause either discontinuity of marketing, or discontinuity of technology, but not both. Really new innovations include combinations of these two extremes. These three definitions of product innovation also indicate a reduction in the degree of innovativeness as follows: radical→ really new → incremental innovation.

An alternative approach to categorize technical change is the scale of technological innovation intensity by Coccia (2005) that measures and classifies technical change according to effects generated by technological innovations on geo-economic space, in analogy with the effects of seismic waves (cf., also Coccia, 2005a).

Pavitt (1984, p. 343ff) proposed a taxonomy of sectoral patterns of technical change based on innovating firms: "(1) supplier dominated; (2) production intensive; (3) science based. They can be explained by sources of technology, requirements of users and possibilities for appropriation. This explanation has implications for our understanding of the sources and directions of technical change, firms' diversification behaviour, the dynamic relationship between technology and industrial structure, and the formation of technological skills and advantages at the level of the firm, the region and the country".

De Marchi (2016, p. 984), instead, endeavors to formulate a classification based on general characteristics of scientific discovery and technological innovation. The features of these two activities can be described with oppositions between pairings of aspects of ''real oppositions'', graphically represented by pairs of semi axes. The first real opposition would be between problems and solutions. The second real opposition adopted is that countering specificity and generality of problems and solutions (cf., Arthur 2009). Since these two oppositions are simultaneously applicable to science and technology, the study categorizes the activities of both research and innovation in a matrix 2×2, where each cell is defined by a pair of semi axes (cf., De Marchi, 2016, pp. 984-985).

In short, the vast literature has suggested many approaches for classification of innovation, though studies described above are not a comprehensive review in these research fields (Clark, 1985; Coccia, 2016; Hargadon, 2003; Nelson and Winter, 1982; Nelson 2008; Rosenberg, 1969; cf. Anadon et al., 2016)[4]. However, studies of technical change have given little systematic attention to the different characteristics of interaction between technologies that can generate coevolution of technological systems and technological change in society. The crux of the study here is to categorize technologies considering their interaction with other technologies, in a broad analogy with the ecology[5]. The suggested interpretation here can provide a theoretical framework to clarify typologies of interactive technologies that support evolutionary pathways of complex systems of technology over time and space. At the same time, we are aware of the vast differences between biological and technological processes (cf., Braun, 1990; Hodgson, 2002; Ziman, 2000).

---

4  See Coccia (2006) for further approaches of classifications of innovation in economics of technical change and management of technology.
5  Ecology is the scientific study of interactions between organisms of the same or different species, and between organisms and their non-living environment (Poulin, 2006). The scope of the ecology is to explain the number and distribution of organisms over time and space and all sorts of interactions.

## Method

In order to lay the foundations for a new taxonomy of technologies here, it is important to clarify the concept of complexity and complex systems. Simon (1962, p. 468) states that: "a complex system [is]... one made up of a large number of parts that interact in a nonsimple way .... complexity frequently takes the form of hierarchy, and .... a hierarchic system ... is composed of interrelated subsystems, each of the latter being, in turn, hierarchic in structure until we reach some lowest level of elementary subsystem." McNerney et al. (2011, p. 9008) argue that: "The technology can be decomposed into $n$ components, each of which interacts with a cluster of $d - 1$ other components" (cf., Arthur, 2009). A characteristic of complex systems is the interaction *between* systems and the interaction *within* systems—i.e., among the parts of those systems. This philosophical background of the architecture of complexity by Simon (1982), shortly described, is important to support theoretically the taxonomy of interactive technologies proposed by the study here.

Taxonomy of interactive technologies is based on following concepts:

- A technology is a complex system that is composed of more than one component or sub-system and a relationship that holds between each component and at least one other element in the set. The technology is selected and adapted in the Environment $E$ with a natural selection operated by market forces and artificial selection operated by human beings to satisfy needs, achieve goals and/or solve problems in human society.

- Interaction between technologies $T1$ and $T2$ or more associated technologies $Ti$ ($i$=1, ..., n) is a reciprocal adaptation between technologies in a complex system $S$ with inter-relationships of information/resources/energy and other physical phenomena to satisfy needs, achieve goals and/or solve problems in human society. $Ti$ is called interactive technology in S.

The proposed taxonomy (TX) here is established to respect the following conditions of (Brandon, 1978, pp. 188-192):

i.  independence: the taxonomy to play its explanatory role cannot be a tautology.

ii.  generality: it must apply to the whole elements of technological change. It must be general and universally applicable throughout the domain of technical and economic change.

iii.  epistemological applicability: TX has to be testable and can be applied to particular cases of systems of technology.

iv.  and empirical correctness: TX must not be false.

Overall, then, the taxonomy suggested here has the goal to categorize and generalize the typologies of interactive technologies and clarify, whenever possible their role in evolutionary pathways of complex systems over time and space.

**A proposed taxonomy and theory of interactive technologies in complex systems**

The basic unit of technology analysis, in the proposed taxonomy and theory, is interactive technologies. In general, technologies do not function as independent systems *per se*, but they depend on other (host) technologies to form a complex system of parts that interact in a non-simple way (*e.g.,* batteries and antennas in mobile devices, etc.; cf., Coccia, 2017). Coccia (2017a) states the theorem of *not* independence of *any* technology that in the long run, the behavior and evolution of *any* technology is *not* independent from the behavior and evolution of the other technologies. In general, technologies are not autonomous systems *per se*, but they form complex systems composed of inclusive and interrelated sub-systems of technologies until the lowest level of technological unit (*cf.,* Simon, 1962, p. 468; Oswalt, 1976; cf., Coccia, 2017, 2017a). To put it differently, technologies can function in ecological niches of other technologies and the interaction between technologies can be an important taxonomic characteristic to categorize technologies that support the coevolution of technological systems (i.e., the evolution of reciprocal adaptations of technologies in a complex system S).

Suppose that the simplest possible case involves only two interactive technologies, *T1* and *T2* in a Complex System *S*(T1, T2); of course, the theory can be generalized for complex systems including many sub-systems of technology, such as *S*(T1, T2, …, T*i*, …T$_N$). Table 1, based on theoretical framework above, categorizes four types of interactive technologies within a complex system S, in a broad analogy with ecology.

**Table 1.** A taxonomy of technologies in complex systems

| Grade | Typology of interactive technology | Examples |
|-------|-----------------------------------|----------|
| 1 | *Technological parasitism* is a relationship between two technologies T1 and T2 in a complex system S where one technology T1 benefits (+) from the interaction with T2, whereas T2 has a negative side (−) from interaction with T1. The interaction between T1 and T2 in mathematical symbols is indicated here (+, −) to represent the benefits (positive or negative) to technologies from interaction in a complex system S(T1,T2). | An example of parasite technology is audio headphones, speakers, software apps, etc. of many electronic devices. These technologies are parasites of different technologies because they can function, if and only if (iff) associated with other technologies. Plus sign (+) indicates the fruitful benefit to parasitic technologies from interaction. In Information and Communication Technologies, host technology decreases its energy from interaction with parasitic technologies, such as electric power of battery; the sign − (minus) here indicates the negative side of interaction for host technology. |
| 2 | *Technological commensalism* is a relationship between two technologies where one technology T1 benefits (+) from the other without affecting it (0). The commensal relation is often between a larger host or master technology and a smaller commensal technology; host or master technology is unmodified from this interaction, whereas commensal technologies may show great structural adaptation consonant with their systems. The interactive technologies (T1, T2) have a relation (+, 0) in a complex system S. 0 (zero) indicates here no benefits from interaction. | An example of commensal technologies is the connection of a single mobile device to a large Wi-Fi network; the connection of an electric appliance to national electricity network; etc. |
| 3 | *Technological mutualism* is a relationship in which each technology benefits from the activity of the other technology. The interaction between T1 and T2 has mutual benefits in S indicated with symbols (+, +). | An example of mutual technologies is the relation between battery and mobile devices, antenna and mobile devices, HD displays and mobile devices, etc. The interaction here generates mutual benefits between technologies (+,+) in S. |
| 4 | *Technological symbiosis* is a long-term interaction between two technologies (T1,T2) that evolve together in a complex system S. The symbiotic technologies have a long-run interaction that generates continuous and mutual benefits and, as a consequence, coevolution of complex systems in which these technologies function and adapt themselves. The interaction between T1 and T2 in S is indicated with (++, ++) to represent benefits of the long-run mutual symbiotic relationship between host and parasitic technologies (coevolution of technological systems). | For instance, symbiotic technologies are the continuous interaction between Bluetooth technology and mobile devices that has improved both technologies and increased their effectiveness and technical performance, such as Bluetooth 2.0 with an Enhanced Data Rate for faster data transfer, Bluetooth 4.0 with low energy to save battery of mobile devices, etc. This technological evolution of Bluetooth technology is associated with new generations of mobile devices –e.g. *i*Phone 6,7,8, etc.– in order to better interact with this and other technologies and generate coevolution of complex systems in which these technologies function (Apple Inc., 2016; Bluetooth, 2017). |

*Note*: +(Plus) is a positive benefit to technology T$i$ from interaction with technology T$j$ in a complex system S ($\forall i=1,...,n; \ \forall j=1,...,m$); −(minus) is a negative benefit to technology T$i$ from interaction with technology T$j$ in S; 0 (zero) indicates a neutral effect from interaction between technologies T$i$ and T$j$ in S; ++ is a strong positive benefit from long-run mutual symbiotic interaction between technologies T$i$ and T$j$ in S (i.e., coevolution of T$i$ and T$j$ in S).

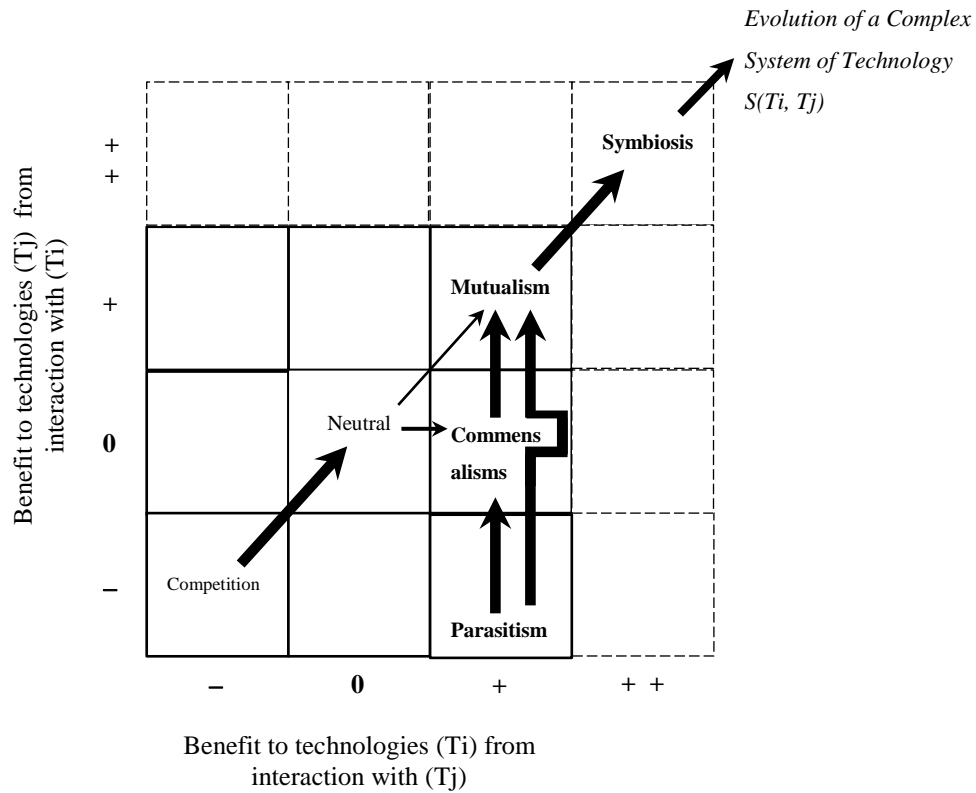

**Figure 1.** Types and evolutionary pathways of interactive technologies in a complex system S. *Note.* The notions of positive, negative and neutral benefit from interaction between technologies *Ti* and *Tj* in S are represented with mathematical symbols +, −, 0 (zero). , ++ is a strong positive benefit from long-run mutual symbiotic interaction between technologies T*i* and T*j* in S (i.e., coevolution of T*i* and T*j* in S). Thick solid arrows indicate the probable evolutionary route of interactive technologies in a complex system S: the possibilities for parasitic technologies to become commensals, mutualists, and symbiotic; thin arrows show other possible evolutionary pathways of technologies T*i* and T*j* during the interaction in a complex system S ( $\forall i=1,...,n; \; \forall j=1,...,m$ ).

In general, parasitism, mutualism, commensalism and symbiosis between technologies do not establish clear cut-offs of these concepts and each relationship represents an end-point of an evolutionary development of interactive technologies in a complex system *S* (cf., Poulin, 2006 for ecological interaction). In particular, parasitism is an interaction that may evolve over time towards commensalism, mutualism and symbiosis to support evolutionary innovations (cf., Price, 1991). The symbiosis is also increasingly recognized as an important selective force behind interdependent coevolution of complex systems (cf., Smith, 1991). In short, the interaction between technologies tends to generate stepwise coevolutionary processes of complex systems (cf., Price, 1991). Figure 1 represents evolutionary pathways of the four typologies of interactive technologies in S (Table 1).

The proposed taxonomy here has the following properties:

1). *Property of increasing interaction of technology in S over time.* Interactive technologies increase the grade of interaction over time directed to evolution of an overall system of technology S along the following evolutionary route: technological parasitism→ commensalism → mutualism → technological symbiosis ⇒ evolution of technology (see, Figure 1).

2) *Property of inclusion of interactive technologies*. Interactive technologies can be of four types (Tab. 1):

TS= Technological Symbiosis; TM= Technological Mutualism; TC=Technological Commensalism; TP= Technological Parasitism.

TS, TM, TC and TP are sets within a complex system S.

The set theory indicates with the symbol $\subset$ a subset. A derived binary relation between two sets is the set inclusion. In particular, interactive technologies of proposed taxonomy have the following property of inclusion in S:

$[(TP \subset TC) \subset TM] \subset TS$ ∎

Overall, then, this taxonomy can systematize the typologies of interactive technologies and predicts their evolutionary pathways that generate stepwise coevolutionary processes within a system of technology S (e.g., devices, new products, etc.).

**Predictions of the taxonomy and theory of interactive technologies**

Technologies are complex systems composed of interrelated technological subsystems until the lowest level of technological unit (cf., Oswalt, 1976). Interaction is proposed here to be one of the mechanisms driving the evolution of technology and a critical taxonomic characteristic for a classification of technology (cf., Coccia, 2017). On the basis of the suggested taxonomy here, it is possible to make some predictions about evolutionary paths of interactive technologies within complex systems S.

a) The short-run behavior and evolution of interactive technologies is approximately independent from the other technologies in S. In particular, the short-run evolution of a specific interactive technology (e.g., parasite technology) is due to advances or mutations in the technology itself.

b) The long-run behavior and evolution of any interactive technologies (i.e., *technological parasitism, commensalism, mutualism and symbiosis*) depends on the behavior and evolution of associated technologies; in particular, the long-run behavior and evolution of any interactive technology is due to interaction with other technologies within and between complex systems.

c) Symbiotic, mutualistic, commensal and parasitic technologies tend to generate a rapid evolution of a complex system of technology S in comparison with complex systems without interactive technologies.

**Discussion of some analytical implications**

The proposed taxonomy and theory here have a number of implications for the analysis of nature, source and evolution of technical change. Some of the most obvious implications, without pretending to be comprehensive are as follows.

*1.1    Contribution to the literature on taxonomy of technical change*

This study contributes to the literature on taxonomy of technical change by detailing the importance of specific typologies of interactive technologies during the evolutionary patterns of technological innovation. Current literature categorizes technical change with *static* characteristic considering objects and/or subjects of technological innovation (Archibugi and Simonetti, 1998; Freeman and Soete, 1987).

In fact, technology can be classified according to: a) the nature of technological innovation-*object*-, such as incremental and radical innovation, product and process innovation, etc. (cf., Freeman and Soete, 1987); b) The sector of activity of innovative firms-*subject*-, such as supplier-dominated, scale-intensive, specialized suppliers and science- based (Pavitt, 1984).

The study here extends this specific literature by identifying typologies of technologies with a *dynamic* characteristic represented by interaction between technologies in complex systems over time. The theoretical framework here categorizes the interaction between technologies in technological parasitism, commensalism, mutualism and symbiosis. These typologies of interactive technologies have specific characteristics that drive the evolutionary pathways of complex systems of technology and technological diversification over time and space. The dynamic characteristic underlying the proposed taxonomy here may also help better understand the linkages between technologies that explain directions of technical development of complex systems of technology. In general, the taxonomy and theory here, borrowing concepts from ecology, it can extend economics of technical change with a new research stream to theorize and categorize interactive technologies that can explain the process through which these technologies become meaningful, and their role for processes of evolution of complex systems of technology.

*1.2   Contribution to the literature on evolution of technology*

This theory here also extends the literature on technological evolution identifying some important but overlooked typologies of technology within the nature of technology (Arthur, 2009; Dosi, 1988). Arthur (2009, pp. 18-19) argues that the evolution in technology is due to combinatorial evolution: "Technologies somehow must come into being as fresh combinations of what already exists". This combination of components and assemblies is organized into systems to some human purpose and has a hierarchical and recursive structure: "technologies … consist of component building blocks that are also technologies, and these consist of subparts that are also technologies, in a repeating (or recurring) pattern" (Arthur, 2009, p. 38). In short, Arthur (2009) claims that a source of change in technology evolution is the combination based on supply of new technologies assembling existing components and on demand for means to fulfill purposes, the need for novel technologies. The suggested taxonomy of technologies here is consistent with this well-established literature by Arthur (2009) as well as with studies that consider structural innovations and systems innovations based on integration of two or more symbiotic technologies (Sahal, 1985). However, the study here extends this research field by detailing how different typologies of technologies interact in complex systems and guide the evolution of technology. One of the most important implications of this work is also that specific interactive technologies, such as symbiotic technologies, can generate fruitful evolutionary routes for complex systems of technology S in evolving industries. Kalogerakis et al. (2010, p. 418) argue that new technology can also be due to 'inventive analogical transfer' from experience of a specific technology in one knowledge field – *source domain* – to other scientific fields – *target domains*[6]. This theory adds to this body of literature a new perspective represented by the interaction between technologies from source domain to other target domains of systems of technology to satisfy needs and/or to solve problems in human society. *Overall, then, the theoretical framework developed here opens the black box of the interaction between technologies that affects, with different types of technologies, the evolutionary pathways of complex systems of technology over time and space.*

**Concluding observations**

Manifold dimensions in the analysis and evolution of technology are hardly known. Researchers should be ready to open the debate regarding the nature and types of interaction between technologies that may explain the evolution of technology and technical change in human society (cf., De Marchi, 2016). Some scholars argue that technologies and technological change display numerous life-like features, suggesting a deep connection with biological evolution (Basalla, 1988; Erwin and Krakauer, 2004; Solé et al., 2011; Wagner and Rosen, 2014). This study extends the broad analogy between technological and biological evolution to more specifically focus on the potential of a taxonomy and theory of interactive technologies in complex systems, but fully acknowledge that interaction between technologies is not a perfect analogy of biological/ecological interaction; of course, there are differences (Ziman, 2000; Jacob, 1977; Solé *et al.,* 2013). For studying technical change, though, the analogy with biology and ecology is a source of inspiration and ideas because it has been studied in such depth and provides a logical structure of scientific inquiry in these research fields. The study here proposes a taxonomy of technology based on four typologies represented by technological parasitism, commensalism, mutualism and symbiosis that can guide evolutionary pathways of technology within and between complex systems. These types of interactive technologies seem to be general driving components for the evolution of new technology across time and space (cf., Smith, 1991; Prince, 1991; Coccia, 2017). The characteristics and dynamics of interactive technologies, described in table 1 and figure 1, are also affected by learning processes and technological capability of firms in markets with rapid change (cf., Teece et al., 1997; Zollo and Winter, 2002).

---

[6] *Cf. also*, Cavallo et al., 2014, 2015; Coccia 2009, 2012, 2012a, 2015; Coccia and Wang, 2015, 2016.

On the basis of arguments presented in this study, the taxonomy here categorizes general typologies of interactive technologies that can explain, whenever possible, some characteristics of the interaction between technologies for the evolution of complex systems of technology and technical change in human society.

In particular, the results here suggest that:

1. Technological parasitism, commensalism, mutualism and symbiosis can help explain aspects of evolutionary pathways of complex systems within technical change in society.

2. Evolution of complex systems of technology may be rapid in the presence of subsystems of technological symbiosis and/or mutualism, rather than technological parasitism and commensalism (*see*, Fig. 1).

Hence, the study here provides an appropriate theoretical framework to classify interactive technologies and explain possible evolutionary pathways of complex systems of technology. Moreover, taxonomy here suggests a general prediction that it may be possible to influence (support) the long-run evolution of technical change by increasing mutual symbiotic interactions between technologies. This finding could aid technology policy and management of technology to design best practices to support technological interaction in complex systems for industrial and economic change, and technological progress of human society. Valverde (2016, p.5) in this context also states that: "Technological progress is associated with more complex human-machine interactions". As a matter of fact, human activity acts as ecosystem engineers able to change social and technological systems (Solé *et al.,* 2013). In short, the study here makes a unique contribution, by showing how technology can be classified in critical typologies considering the concept of interaction between technologies. This idea of a "taxonomy of interactive technologies" suggested in the study here is adequate in some cases but less in others because of the vast diversity of technologies and their interaction in complex systems and environments. Nevertheless, the analogy keeps its validity in classifying and explaining general interaction and coevolution of technology in complex systems. The taxonomy here also suggests some properties of interactive technologies that are a reasonable starting point for understanding the universal features of the technology and coevolution of complex systems of technology that leads to technical change and progress in society, though the model here of course cannot predict any given characteristics of technologies with precision.

These typologies of interactive technologies can create theoretically, methodological and empirical challenges. In particular, scholars studying technology and technological evolution might have to take the interaction between technologies into account and begin data collection to explain with comprehensive model the role of interactive technologies for the emergence and evolution of technological paradigms and trajectories (Nelson and Winter, 1982; Dosi, 1988). Future efforts in this research stream will be directed to provide empirical evidence of the interaction between technologies in complex systems to better classify and evaluate their role during the process of evolution of new technology and, in general, of technical change. Other directions for the future of this research topic, which is not a studied field, are: firstly, the proposed taxonomy needs to be tested on the basis of complete coverage of different technologies belonging to many sectors; secondly, this taxonomy needs to be extended; thirdly, the taxonomy may be studied to provide a variety of uses for designing best-practices of innovation policy and management of technology; finally, the taxonomy and the theory here may be studied to shed light on a number of important aspects of technical change, such as new types, directions and routes of interactive technologies in different industries, accumulation of technological skills and dynamic capabilities of firms from interaction between technologies in markets with rapid change, emerging technologies from interactive technologies, etc. (cf., Teece et al., 1997).

Overall, then, this taxonomy may support a better understanding of the role played by interactive technologies in evolutionary patterns of technological innovation and in general social and technical change. In addition, given the variety of technologies in current patterns of technological change, the taxonomy here can support a generalization and systematization of typologies of interactive technologies during the evolution of technology. Although, we know that other things are often not equal over time and space in the domain of technology.

To conclude, the proposed taxonomy here based on the ecology-like interaction between technologies—may lay the foundation for development of more sophisticated concepts and theoretical frameworks in economics of technical change. In particular, this study constitutes an initial significant step in categorizing technologies considering the interaction between technologies in complex systems and evolution of technology inexorably interlinked. However, identifying generalizable taxonomy and theory is a non-trivial exercise. Wright (1997, p. 1562) properly claims that: "In the world of technological change, bounded rationality is the rule."